\documentclass[superscriptaddress,showpacs,twocolumn,prd]{revtex4}
\usepackage{graphicx}

\usepackage{amsmath}
\newcommand{\be}{\begin{equation}}
\newcommand{\ee}{\end{equation}}
\newcommand{\ba}{\begin{eqnarray}}
\newcommand{\ea}{\end{eqnarray}}

\newcommand{\Q}{{\cal Q}}

\newcommand{\Tr}{\mathrm{Tr}}
\newcommand{\Str}{\mathrm{Str}}
\newcommand{\diag}{\mathrm{diag}}
\newcommand{\nn}{\nonumber\\}
\newcommand{\lgl}{\langle}
\newcommand{\rgl}{\rangle}

\begin{document}

\onecolumngrid
\thispagestyle{empty}
\begin{flushright}
{\large 
LU TP 06-38\\
\large hep-lat/0610127\\[0.1cm]
\large October 2006}
\end{flushright}
\vskip2cm
\begin{center}
{\Large\bf
Electromagnetic Corrections in Partially\\[4mm]
Quenched Chiral Perturbation Theory}

\vfill

{\large \bf Johan Bijnens$^a$ and Niclas Danielsson$^{a,b}$}\\[1cm]
{$^a$Department of Theoretical Physics, Lund University, \\
S\"olvegatan 14A, SE - 223 62 Lund, Sweden\\[1cm]
$^b$Division of Mathematical Physics, 
Lund Institute of Technology, Lund University, \\ 
Box 118, SE - 221 00 Lund, Sweden
}

\vfill

{\large\bf Abstract}

\vskip1cm

\parbox{14cm}{\large
We introduce photons in Partially Quenched Chiral Perturbation Theory
and calculate the resulting electromagnetic loop-corrections
at NLO for the charged meson masses and decay constants. We also present a
numerical analysis to indicate the size of the different corrections.
We show that several phenomenologically relevant quantities can be calculated
consistently with photons which couple only to the valence quarks, allowing the
use of gluon configurations produced without dynamical photons.
}

\vskip3cm

\end{center}
\vskip3cm
\twocolumngrid
\setcounter{page}{0}

\title{Electromagnetic Corrections in Partially Quenched Chiral
Perturbation Theory}

\author{Johan Bijnens}
\affiliation{Department of Theoretical Physics, Lund University,\\
S\"olvegatan 14A, SE - 223 62 Lund, Sweden}
\author{Niclas Danielsson}
\affiliation{Department of Theoretical Physics, Lund University,\\
S\"olvegatan 14A, SE - 223 62 Lund, Sweden}
\affiliation{Division of Mathematical Physics, Lund Institute of 
Technology, Lund University,\\ 
Box 118, SE - 221 00 Lund, Sweden}


\begin{abstract} 
We introduce photons in Partially Quenched Chiral Perturbation Theory
and calculate the resulting electromagnetic loop-corrections
at NLO for the charged meson masses and decay constants. We also present a
numerical analysis to indicate the size of the different corrections.
We show that several phenomenologically relevant quantities can be calculated
consistently with photons which couple only to the valence quarks, allowing the
use of gluon configurations produced without dynamical photons.

\end{abstract}

\maketitle

\section{Introduction}
\label{intro}
This is the sixth paper in a series of studies of Partial Quenching (PQ) in
Chiral Perturbation Theory ($\chi$PT)~\cite{Weinberg,GL1,GL2}, 
where we now investigate the effects
of including electromagnetic loop corrections in the theory. The motivation
for using partial quenching in $\chi$PT (and indeed for studying $\chi$PT
itself) comes from the fact that even though quantum chromodynamics (QCD) 
over time has become the generally
accepted theory of the strong interaction, it has still proven difficult 
to use this
theory to derive
low-energy hadronic observables such as masses and decay constants. 

An
alternative approach is to use Lattice QCD simulations to this
end. Computational limitations have however hindered such simulations for light
particles since they can propagate over large distances, requiring very large
lattice sizes. Because of this, most simulations have so far been performed 
with heavier quark masses than those of the physical world.

On the other hand,  Chiral Perturbation Theory provides a theoretically
correct description of the low-energy properties of QCD, and can be used to
extrapolate the results of lattice simulations down to the masses of the 
physical regime of QCD. In
particular, one can use lattice simulations to determine the low-energy
constants of $\chi$PT by fitting $\chi$PT calculations to corresponding
lattice simulations, thereby getting estimates of hadronic low-energy
observables.

One problem with this approach is that reliable predictions from 
$\chi$PT require that one keeps the quark masses
fairly small, and so far it has proven difficult to reach the chiral
regime in the lattice simulations. However, progress is being made on
this front.
A complementary approach is
therefore to use {\it{partial quenching}} where one introduces a 
separate quark mass for
the calculations of closed quark loops, referred to as sea quarks, compared to
the quark lines which are connected to external sources, referred to as
valence quarks.
This has 
the advantage over full QCD calculations that results with more
values of the valence quark masses can be obtained with a smaller number
of values of sea quark masses, which is useful since varying the latter is 
computationally more expensive.

Unquenched QCD may be recovered from partially quenched QCD (PQQCD) by taking 
the limit of equal sea and valence quark masses, and therefore it follows 
that QCD and PQQCD
are continuously connected by the variation of sea-quark masses. This means
that, in contrast to a fully quenched theory where the effects of the closed
quark loops are neglected altogether, one can relate partially quenched QCD
simulations to the unquenched physical observables of the real world.

Chiral Perturbation Theory has also been extended to include both
quenching and
partial quenching~\cite{Morel,BG1,SharpeA,BG2}. The formulation of Partially
Quenched $\chi$PT (PQ$\chi$PT) is such that the dependence on the
quark masses is explicit, and thus the limit of equal sea and valence quark
masses can also be considered for PQ$\chi$PT. This allows for determination 
of the physically relevant LECs of $\chi$PT by fits of partially 
quenched $\chi$PT (PQ$\chi$PT) to partially quenched lattice simulations 
(PQQCD), see e.g. the discussion in~\cite{Sharpe1}. In
particular, the LECs of $\chi$PT, which are of physical significance, can be
obtained directly from those of PQ$\chi$PT. More detailed discussions of this
and references to earlier work can be found in the papers of Sharpe and 
Shoresh~\cite{Sharpe1,Sharpe2}. 
The calculations in this paper have been performed in
three-flavor PQ$\chi$PT without
the $\Phi_0$~\cite{Sharpe2} degree of freedom.
In our earlier work with Timo L\"ahde~\cite{BDL,BL1,BL2,BDL2,BD1} we have
calculated masses and decay constants for the charged, or off-diagonal, mesons
to next-to-next-to-leading-order in PQ$\chi$PT.
However, in order to compare with the experimental values at high precision
one needs to take electromagnetic effects into account as well.

Electromagnetic corrections in $\chi$PT have a long history. The lowest order
(LO)
was done by Dashen~\cite{Dashen}. The first corrections to this
were worked out in Ref.~\cite{Langacker}. That large corrections might appear
was pointed out in Refs.~\cite{DHW,BijnensDashen,Maltman}, where these
corrections appear both from chiral logarithms and effects persisting at
large $N_c$. The work in pure $\chi$PT was started by Urech\cite{Urech1}
and by Neufeld and Rupertsberger~\cite{Neufeld}. The NLO expressions for
the masses were calculated in both papers and the decay constants in 
the second.
More recent work on estimating the relevant LECs can be found in
Refs.~\cite{BP,Moussallam1,Donoghue,Moussallam2,Moussallam3,Pinzke}.
Note that there are several subtleties involved in defining electromagnetic
corrections at low energies as discussed in Refs.~\cite{BP,Moussallam1,Gasser}
with increasing levels of detail. There are also first exploratory lattice QCD
calculations~\cite{Lattice1,Lattice2,Lattice3,Lattice4}.

In this paper we present the extension of PQ$\chi$PT to include dynamical
photons to NLO. In addition we calculate
the NLO expressions for the masses of the charged, or offdiagonal, mesons
with virtual photon-loop corrections for all possible degrees of degeneracy in
the valence- and sea-quark masses. We do the same for the decay
constants,
i.e. we determine the ${\cal O}(e^2)$ and ${\cal O}(e^2 p^2)$ corrections
in PQ$\chi$PT to all these quantities.

We point out that the two phenomenologically relevant quantities
$\Delta M^2$ and $\Delta F$ for differences of masses and decay constants,
can be determined from partially quenched lattice calculations where the
photons are only coupled to the valence quarks. This has the important
consequence that these quantities can be calculated in lattice QCD with gluon
configurations generated without dynamical photons.

The paper is organised as follows. First we present the technical background
and notation already present in the earlier work~\cite{BDL,BL1,BL2,BDL2,BD1}
in Sect.~\ref{background}. We also present the results for the needed loop
integrals there. The extension to dynamical photons is discussed
in Sect.~\ref{photons}. Here we give the Lagrangians needed, as well
as the subtractions needed to obtain finite results. The analytical expressions
for the masses and decay constants are given in Sect.~\ref{Analytical} and
discussed in Sect.~\ref{analytical}. Some illustrative numerical results
are give in Sect.~\ref{numerical} and we recapitulate the main conclusions
in Sect.~\ref{conclusions}.

\section{PQ$\chi$PT, technical overview}
\label{background}

Here we give a short overview of the technical aspects of PQ$\chi$PT. 
A more thorough discussion of the technical aspects of
PQ$\chi$PT calculations to NLO (without photons) can be found in 
Refs.~\cite{Sharpe1,Sharpe2}.
Our earlier papers, in particular~\cite{BDL2},  also contain
overviews of the NLO technical and notational details, but the focus there is
mainly on the NNLO aspects relevant for those papers. Lectures on standard 
$\chi$PT can be found in~\cite{CHPTlectures}.

The mechanism which gives different masses to sea quarks and valence quarks in
PQ$\chi$PT is introduced by adding explicit sea quarks, as well as unphysical
bosonic ghost quarks. The bosonic quarks are needed to cancel all effects of closed 
loop
contributions from valence quarks. This cancellation happens if the masses
of the bosonic quarks are set equal to the masses of the valence quarks. The
symmetry group of PQ$\chi$PT is essentially given by the graded group 
\begin{equation}
\label{Gchiral}
G = SU(n_\mathrm{val}+n_\mathrm{sea} | n_\mathrm{val})_L
\times SU(n_\mathrm{val}+n_\mathrm{sea} | n_\mathrm{val})_R\,.
\end{equation}
where $n_\mathrm{val}$ denotes the number of valence quarks and
$n_\mathrm{sea}$ the number of sea quarks in the theory. The number of bosonic
quarks is by necessity equal to the number of valence quarks.
The PQ analog to the field matrix $U$ in ordinary $\chi$PT is given by
\begin{equation}
U \equiv \exp\left(i\sqrt{2}\,\Phi/ F_0 \right)\,.
\end{equation}
The matrix $\Phi$ is now a graded matrix, which in terms of a sub-matrix notation for the flavor 
structure can be written as
\begin{equation}
\label{SUSY_FieldMatrix}
\Phi =
\left(\begin{array}{ccc}
\Big[\;\;q_V\bar q_V\;\;\Big] & 
\Big[\;\;q_V\bar q_S\;\;\Big] &
\Big[\;\;q_V\bar q_B\;\;\Big] \\ \\ 
\Big[\;\;q_S\bar q_V\;\;\Big] &
\Big[\;\;q_S\bar q_S\;\;\Big] & 
\Big[\;\;q_S\bar q_B\;\;\Big]\\ \\
\Big[\;\;q_B\bar q_V\;\;\Big] & 
\Big[\;\;q_B\bar q_S\;\;\Big] &
\Big[\;\;q_B\bar q_B\;\;\Big]
\end{array}\right)\,.
\end{equation}
The brackets denote matrices of the form
\begin{equation}
\label{SubMatrix}
q_a\bar q_b =
\left(\begin{array}{ccc}
u_a \bar u_b & u_a \bar d_b & u_a \bar s_b \\ d_a \bar u_b &
d_a \bar d_b & d_a \bar s_b \\ s_a \bar u_b & s_a \bar d_b & s_a \bar s_b    
\end{array}\right)\,,
\end{equation}
where we have used three quark flavors $u$, $d$ and $s$ and 
the labels $V,S$ and $B$ in the sub-matrices stand for valence, 
sea and bosonic 
quarks, respectively. In general, the size of each sub-matrix depends on the exact 
number of quark flavors used, but for this paper all blocks in Eq.~(\ref{SUSY_FieldMatrix}) are $3\times 3$ blocks 

The quarks $q_V$, $q_S$ and their respective antiquarks are fermions, 
while the quarks $q_B$ and their antiquarks are bosons. Each sub-matrix in 
Eq.~(\ref{SUSY_FieldMatrix})
therefore consists of either fermionic or bosonic fields 
only. This construction means that $\Phi$ satisfies the usual rules for
cyclicity under trace and determinant products, provided that we perform the
corresponding sypersymmetric operations instead. For the trace, we must
instead take the supertrace, defined by
\begin{equation}
\Str \left(\begin{array}{cc} A & B \\ C & D \end{array}\right)
=\Tr\,A - \Tr\,D\,,
\end{equation} 
where $A, D$,denote block matrices with commuting elements and $B, C$ denote
block matrices with anticommuting (fermionic) elements. 

This also has the very useful consequence that the Lagrangian structure
of PQ$\chi$PT is the same as for 
$n$-flavor $\chi$PT, 
provided that the traces of 
matrix products in those Lagrangians are replaced by supertraces.
A detailed discussion about the Lagrangians and LECs for the different versions
of $\chi$PT and PQ$\chi$PT without the $\Phi_0$ can be found in~\cite{BDL2}.
This correspondence between $n$-flavor $\chi$PT and PQ$\chi$PT also holds for
the divergence structure when replacing $n$ with the number of sea-quarks.
The same also holds for the extension to electromagnetism in the next section.

In the following, the different quark masses are 
identified by the flavor indices 
$i=1,\ldots,9$, rather than by the indices $u,d,s$ and $V,S,B$ of 
Eqs.~(\ref{SubMatrix}) and~(\ref{SUSY_FieldMatrix}). The results are 
expressed in terms of the quark masses $m_q$ via the quantities 
$\chi_i=2B_0\,m_{qi}$, where $B_0$ is related to the quark-anti-quark
vacuum expectation value in the chiral limit.
Thus $\chi_1,\chi_2,\chi_3$, belong to the 
valence sector, $\chi_4,\chi_5,\chi_6$ to the sea sector, and 
$\chi_7,\chi_8,\chi_9$ to the ghost sector. 
Since we set the quark masses of the
ghost sector equal to the quark masses of the valence sector, the masses 
$\chi_7,\chi_8,\chi_9$ do not appear in the final analytical results.

\subsection{Loop Integrals and Notation}

The expressions for the NLO masses and decay constants of the charged 
pseudoscalar mesons depend on several loop integrals. The renormalized
contributions from these integrals are written in terms 
of the functions
\begin{eqnarray}
\bar A(\chi) &=& -\pi_{16}\, \chi \log(\chi/\mu^2), \nonumber \\
\bar B(\chi,\chi;0) &=& -\pi_{16}\left(1+\log(\chi/\mu^2) \right),\nonumber \\
\bar B(\chi_\gamma,\chi;\chi) &=& 
\pi_{16}\left(1-\log(\chi/\mu^2) \right)+{\cal O}(\chi_\gamma),\nonumber \\
\bar B'(\chi_\gamma,\chi;\chi) &=& 
\frac{\pi_{16}}{\chi}\left(1+\frac{1}{2}\log(\chi/\chi_\gamma) \right)+{\cal O}(\chi_\gamma),\nonumber \\
\bar B_1(\chi_{\gamma},\chi;\chi) &=&
-\frac{\pi_{16}}{2}\log(\chi/\mu^2)+{\cal O}(\chi_\gamma),\nonumber \\
\bar B_1'(\chi_{\gamma},\chi;\chi) &=&
\pi_{16}/(2\chi)+{\cal O}(\chi_\gamma),
\end{eqnarray}  
where $\mu$ denotes the renormalization scale and $\pi_{16} = 1/(16 
\pi^2)$. The argument $\chi_{\gamma}$ is a small photon mass introduced to
regulate infrared divergences.  The prime indicates a derivative with respect
to the momentum squared in the loop integral.

The quantities $d_{\mathrm{val}}$ and $d_{\mathrm{sea}}$ are used to indicate
the number of nondegenerate quark masses in the valence sector and the sea
sector respectively. For $d_{\mathrm{val}}=1$, one has $\chi_1=\chi_2=\chi_4$,
while $d_{\mathrm{val}}=2$ means
$\chi_1=\chi_2\ne\chi_4$. $d_{\mathrm{val}}=3$ is not needed for this
paper. Similarly, $d_{\mathrm{sea}}=1$ means $\chi_4=\chi_5=\chi_6$,
$d_{\mathrm{sea}}=2$ means $\chi_4=\chi_5\ne\chi_6$ and for
$d_{\mathrm{sea}}=3$ all the sea quark masses are nondegenerate, such 
that $\chi_4\ne\chi_5\ne\chi_6$.  

The lowest order 
neutral pion and eta meson masses in the sea quark sector show up at several
places in the
analytical results. They are 
denoted by $\chi_\pi$ and $\chi_\eta$ and are given by the relations
\begin{eqnarray}
\label{neutral_sea_masses}
\chi_\pi+\chi_\eta &=& \frac{2}{3}\left(\chi_4+\chi_5+\chi_6\right),
\nonumber\\
\chi_\pi \chi_\eta &=& \frac{1}{3}
\left(\chi_4\chi_5+\chi_5\chi_6+\chi_4\chi_6\right),
\end{eqnarray}
which have no polynomial solution for $d_{\mathrm{sea}}=3$, but for $d_{\mathrm{sea}}=2$ one 
has $\chi_\pi=\chi_4$ and $\chi_\eta=(\chi_4+2\chi_6)/3$. For
$d_{\mathrm{sea}}=1$ this simplifies further into $\chi_\pi=\chi_\eta=\chi_4$.
The neutral meson propagators in PQ$\chi$PT generate certain reoccuring
combinations of the sea and valence quark masses. An overview of this can be
found in~\cite{BDL2}. The relevant quark-mass combinations for this paper can be expressed in terms of the general quantities
$R^z_{a\ldots b}$ defined by  
\begin{eqnarray}
R^z_{ab} &=& \chi_a - \chi_b, \nonumber \\
R^z_{abc} &=& \frac{\chi_a - \chi_b}{\chi_a - \chi_c}, \nonumber \\ 
R^z_{abcd} &=& \frac{(\chi_a - \chi_b)(\chi_a - \chi_c)}
{\chi_a - \chi_d}, \nonumber \\
R^z_{abcdefg} &=& \frac{(\chi_a - \chi_b)(\chi_a - \chi_c)(\chi_a - \chi_d)}
{(\chi_a - \chi_e)(\chi_a - \chi_f)(\chi_a - \chi_g)},
\label{RSfunc}
\end{eqnarray}
and so on. For the case of $d_{\mathrm{sea}} = 3$, the needed combinations are
\begin{eqnarray}
R_{jkl}^{i} &=& R^z_{i456jkl}, \nonumber \\
R_{i}^{d} &=& R^z_{i456\pi\eta}, \nonumber \\
R_{i}^{c} &=& R^i_{4\pi\eta} + R^i_{5\pi\eta} + R^i_{6\pi\eta}
          - R^i_{\pi\eta\eta} - R^i_{\pi\pi\eta}\nonumber \\
R^v_{ijkl} &=& R^i_{jkk} + R^i_{jll} - 2 R^i_{jkl}\,.
\end{eqnarray}
For the case of 
$d_{\mathrm{sea}} = 2$, corresponding combinations are
\begin{eqnarray}
R_{jk}^{i} &=& R^z_{i46jk}, \nonumber \\
R_{i}^{d} &=& R^z_{i46\eta},\nonumber \\
R_{i}^{c} &=& R^i_{4\eta} + R^i_{6\eta} - R^i_{\eta\eta},
\end{eqnarray}
and for $d_{\mathrm{sea}} = 1$, one has
\begin{eqnarray}
R_{j}^{i} &=& R^z_{i4j}, \nonumber \\
R_{i}^{d} &=& R^z_{i4}.\nonumber \\
R_{i}^{c} &=& 1.
\end{eqnarray}

For certain sums and differences of 
quark-masses, or electric quark charges, we introduce shorthand notation 
given by
\ba
\bar \chi_1=&=& \frac{1}{3}\sum_{i\,=\,4,5,6} \chi_i,\nonumber \\
q_{ij} &=&q_i-q_j,\nonumber \\
\overline Q_2 &=& \frac{1}{3}\sum_{i\,=\,4,5,6} q_i^2 .
\ea
The quark charges are expressed in terms of the unit charge $e$.
$q_{ij}$ is the charge of a meson with flavor quantum numbers
of quarks $q_i\bar q_j$.

The summation conventions from Ref.~\cite{BDL2} have as well been implemented 
where possible. In short, they are as follows: 

\begin{itemize}
\item If the index 
$s$ is present, the entire term is to be summed 
over all sea quark indices. 

\item If the index $q$ is present in a term, there will
always be an index $p$ and the resulting sum is over the pairs
of valence indices.  If 
only $p$ is present, the sum is just over the valence indices.
If we  choose valence quarks of type 1 and 3, this becomes
summing over
$(p,q) = (1,3)$ and $(p,q) = (3,1)$ or if 
only $p$ is present, the sum is over $p=1$ and $p=3$.

\item If the index $m$ is present, there 
will always be an index $n$ and the corresponding sum is over the pairs 
$(m,n) = (\pi,\eta)$ and $(m,n) = (\eta,\pi)$. If only the index $n$ 
is present, 
then the term is to be summed over the $\chi_\pi$ and $\chi_\eta$ 
masses.
\end{itemize}

\section{Virtual photons}
\label{photons}

In $\chi$PT, photons are included as external vector fields through a charge
matrix Q for the three light quarks and through the covariant derivative
$D_\mu$~\cite{Urech1}. Introducing photons in PQ$\chi$PT is completely
analogous, provided that traces are replaced by supertraces, in the following
denoted by
$\langle \cdots \rangle$, and that we use
the $n$-flavor expressions for the Lagrangians. The
  covariant derivative includes the photon field through
\be
D_\mu U= \partial_\mu U-ir_\mu U + iUl_\mu,
\ee
with
\ba
r_\mu &=& v_\mu+eQ_R A_\mu+a_\mu\nonumber\\
l_\mu &=& v_\mu+eQ_L A_\mu-a_\mu.
\ea 
For the meson masses, we set $v_\mu=a_\mu=0$ and for the decay constants
$v_\mu=0$. The charge matrix $Q_{\rm L, \rm R}$ is the natural
generalisation of the $SU(3)$ charge
matrix in~\cite{Urech1}. $e$ is the absolute value of the electron charge
for physical quantities but is a free parameter in the lattice calculations.
The $\rm L/ \rm R$ notation refers to the symmetry properties assigned to the
$Q$'s during the construction of the allowed Lagrangian terms~\cite{Urech1},
but in the
usual physical picture the charge matrix is a constant matrix, and one has
$Q_{\rm L}=Q_{\rm R}=Q$, where $Q$ is a diagonal matrix given by
\be
Q =\diag (q_1,\dots,q_9).
\ee 
However, for the notation used below, the distinction
between the two types is still needed. 
Furthermore, one would normally set $q_1=2/3$ and $q_2=q_3=-1/3$ to agree
with ordinary
photon-included $\chi$PT for the real world, but for greater generality we
have kept the $q_i$'s free in the analytical results in this paper. In ordinary
$\chi$PT one requires the charge matrix to be traceless. For
PQ$\chi$PT, $Q$ is a
graded matrix where we set $q_7=q_1, q_8=q_2$, and $q_9=q_3$.
This, together with the earlier requirement on the masses insures that
the closed valence quark loops with photons coupled to them also cancel against
the corresponding ghost quark loops. 
Therefore the
PQ$\chi$PT requirement $\langle Q\rangle =0$ becomes
\be
q_4+q_5+q_6=0.
\ee
Finally, the quark masses are present through the matrix
\be
\chi=\diag (\chi_1,\dots,\chi_9), \quad \chi_i=2B_0m_{q_i}.
\ee   
 For convenience, the
Lagrangians below will be written in terms of the field matrix
\begin{equation}
u\equiv\exp\left(i\Phi/(\sqrt{2} F_0)\right),
\end{equation}
which is related to $U$ through $u=\sqrt{U}$. We also introduce 
the quantities
\begin{eqnarray}
u_\mu &=& i\left\{
u^\dagger(\partial_\mu-i r_\mu)\,u -
u\,(\partial_\mu-i l_\mu)\,u^\dagger\right\},
\nonumber \\
\chi_\pm &=& u^\dagger\chi\,u^\dagger\pm u\,\chi^\dagger\,u,
\nonumber \\
f_\pm^{\mu\nu} &=& u\,F_L^{\mu\nu}\,u^\dagger\pm 
u^\dagger F_R^{\mu\nu}\,u\nonumber \\
\Q_{\rm L} &=& uQ_{\rm L} u^\dagger\nonumber \\
\Q_{\rm R} &=& u^\dagger Q_{\rm R} u\nonumber \\
\hat \nabla_\mu \Q_{\rm L} &=& uD_\mu Q_{\rm L} u^\dagger\nonumber \\
\hat \nabla_\mu \Q_{\rm R} &=& u^\dagger D_\mu Q_{\rm R} u,
\label{uquant}
\end{eqnarray}
where $F_L$ and $F_R$ denote the field strengths of the external fields 
$l$ and $r$, such that 
\ba
F_L^{\mu\nu} &=& 
\partial^\mu l^\nu-\partial^\nu l^\mu-i\left[l^\mu,l^\nu\right],\nonumber\\ 
F_R^{\mu\nu} &=& 
\partial^\mu r^\nu-\partial^\nu r^\mu-i\left[r^\mu,r^\nu\right].
\ea
and the covariant derivatives of $Q_L,Q_R$ are defined by 
\ba
D^\mu Q_L &=& 
\partial^\mu Q_L-i\left[l^{\mu},Q_L \right],\nonumber\\ 
D^\mu Q_R &=& 
\partial^\mu Q_R-i\left[r^{\mu},Q_R \right].
\ea
The quantities in Eq.~(\ref{uquant}) have a well-defined and 
simpler (as elaborated in~\cite{BDL2,BCE1}) behaviour 
under the symmetry transformations
needed for the construction of the Lagrangians. 
In this notation, the lowest order Lagrangian has the form
\ba
\label{lagLO}
{\cal L}_2 = &-&\frac{1}{4}F_{\mu\nu}F^{\mu\nu}-\frac{1}{2}\lambda
(\partial_\mu A^\mu)^2\nonumber\\
&+&\frac{ F^2_0}{4} \langle u^\mu u_\mu + \chi_+\rangle\nonumber\\
&+&e^2C\langle \Q_{\rm L}\Q_{\rm R}\rangle,
\ea
where $F_{\mu\nu}$ is the field strength tensor of the photon field $A_\mu$,
with $F_{\mu\nu}=\partial_\mu A_\nu -\partial_\nu A_\mu$.
Furthermore, $\lambda$ is the gauge fixing parameter, here set to $\lambda=1$, and $e$
is the electric unit charge. We will also use the notation $Z_E = C/F_0^4$.
The lowest order Lagrangian contains terms of ${\cal O}(p^2)$
and ${\cal O}(e^2)$.

For ${\cal L}_4$, the
result is as well analogous to~\cite{Urech1}, except that the terms presented
there are for $SU(3)$. For the $n$-flavor case needed in PQ$\chi$PT, one has
two additional LECs due to the fact that the Cayley-Hamilton relations needed
for the derivation of ${\cal L}_4^{(Q)}$ only are true for the $SU(3)$
case. We split the NLO Lagrangian into the purely strong part of
${\cal O}(p^4)$ and
the part including electromagnetic interactions up to ${\cal O}(e^2p^2)$, and
thus write
\be
{\cal L}_4 ={\cal L}_{S4}+{\cal L}_{S2E2}. 
\ee  
The strong part is given by
\begin{eqnarray}
\label{L4strong}
{\cal L}_{S4} &=& \sum_{i=0}^{12} L_i X_i + \mbox{contact terms}
\nn  
&=& L_0\,\lgl u^\mu u^\nu u_\mu u_\nu \rgl 
+L_1\,\lgl  u^\mu u_\mu \rgl^2 
+L_2\,\lgl u^\mu u^\nu \rgl \lgl u_\mu u_\nu \rgl
\nn
&+&L_3\,\lgl (u^\mu u_\mu)^2 \rgl
+ L_4\,\lgl u^\mu u_\mu \rgl \lgl \chi_+\rgl 
+ L_5\,\lgl u^\mu u_\mu \chi_+ \rgl 
\nn
&+& L_6\,\lgl \chi_+ \rgl^2 
+ L_7\,\lgl \chi_- \rgl^2
+ \frac{L_8}{2}\,\lgl \chi_+^2 + \chi_-^2 \rgl 
\nn
&-& iL_9\,\lgl f_+^{\mu\nu} u_\mu u_\nu \rgl 
+ \frac{L_{10}}{4}\,\lgl f_+^2 - f_-^2 \rgl 
\nn
&+& H_1\,\lgl F_L^2+F_R^2\rgl 
+ H_2\,\lgl\chi\chi^\dagger\rgl,
\end{eqnarray}
where the $L_i$ and $H_i$ are the partially quenched LECs for the case
with three sea quark flavors. 

The electromagnetic part to ${\cal O}(e^2p^2)$ is
\be
\label{L4em}
{\cal L}_{S2E2} = e^2F_0^2\left(\sum_{i=1}^{14} K_{i}^E Q^s_i+K^E_{18} Q^s_{18}+K^E_{19} Q^s_{19}\right).
\nn  
\ee
with
\ba
Q^{s}_1 &=& \frac{1}{2} \; \langle \Q_{\rm L}^2 +
\Q_{\rm R}^2\rangle \; \langle u_\mu u^\mu\rangle \nonumber\\
Q^{s}_2 &=& \langle \Q_{\rm L} \Q_{\rm R}\rangle 
\; \langle u_\mu u^\mu \rangle \nonumber\\
Q^{s}_3 &=& -\langle \Q_{\rm L} u_\mu\rangle 
\; \langle \Q_{\rm L} u^\mu
\rangle - \langle \Q_{\rm R} u_\mu\rangle 
\; \langle \Q_{\rm R} u^\mu\rangle  \nonumber\\
Q^{s}_4 &=& \langle \Q_{\rm L} u_\mu\rangle 
\; \langle \Q_{\rm R} u^\mu \rangle \nonumber\\
Q^{s}_5 &=& \langle(\Q_{\rm L}^2 + \Q_{\rm R}^2)\, 
u_\mu u^\mu\rangle \nonumber\\ 
Q^{s}_6 &=& \langle (\Q_{\rm L} \Q_{\rm R} + 
\Q_{\rm R} \Q_{\rm L})\, u_\mu u^\mu\rangle \nonumber\\
Q^{s}_7 &=& \frac{1}{2} \; \langle \Q_{\rm L}^2 
+ \Q_{\rm R}^2\rangle \; \langle \chi_+\rangle \nonumber\\ 
Q^{s}_8 &=& \langle \Q_{\rm L} \Q_{\rm R}\rangle 
\; \langle \chi_+\rangle \nonumber\\
Q^{s}_9 &=& \langle (\Q_{\rm L}^2 + \Q_{\rm R}^2) \chi_+\rangle \nonumber\\
Q^{s}_{10} &=& \langle(\Q_{\rm L} \Q_{\rm R} 
+ \Q_{\rm R} \Q_{\rm L}) \chi_+\rangle \nonumber\\
Q^{s}_{11} &=& \langle(\Q_{\rm R} \Q_{\rm L} 
- \Q_{\rm L} \Q_{\rm R}) \chi_-\rangle \nonumber\\
Q^{s}_{12} &=& i\langle \left[\hat \nabla_\mu \Q_{\rm R},
               \Q_{\rm R}\right] u^\mu-\left[\hat \nabla_\mu \Q_{\rm L},
                \Q_{\rm L}\right] u^\mu  \rangle \nonumber\\
Q^{s}_{13} &=& \langle \hat \nabla_\mu \Q_{\rm L} \hat \nabla^\mu 
              \Q_{\rm R} \rangle \nonumber\\
Q^{s}_{14} &=& \langle  \hat \nabla_\mu \Q_{\rm L} \hat \nabla^\mu 
               \Q_{\rm L}+ \hat \nabla_\mu \Q_{\rm R} \hat \nabla^\mu 
               \Q_{\rm R} \rangle \nonumber\\
Q^{s}_{18} &=& \langle  \Q_{\rm L} u_\mu  \Q_{\rm L} u^\mu
               +\Q_{\rm R} u_\mu  \Q_{\rm R} u^\mu \rangle \nonumber\\
Q^{s}_{19} &=& \langle \Q_{\rm L} u_\mu \Q_{\rm R} 
               u^\mu \rangle \,. 
\ea
For the $K^E_{i}$ and $Q^{s}_{i}$, we follow the numbering convention
introduced
by Urech~\cite{Urech1}, but $K^E_{15},K^E_{16},K^E_{17}$ are of 
${\cal O}(e^4)$ and
are not needed here. The new terms, needed for the partially quenched
case are thus
named $K^E_{18}$ and $K^E_{19}$.

The relation with constants $K_i$ of Urech when setting valence
and sea-quark masses
equal is
\begin{eqnarray}
  \label{eq:relUrech}
K_1 &=&  K^E_1+K^E_{18}\,,
\nonumber\\
K_2 &=&  K^E_2+\frac{1}{2}K^E_{19}\,,
\nonumber\\
K_3 &=&  K^E_1-K^E_{18}\,,
\nonumber\\
K_4 &=&  K^E_4+K^E_{19}\,,
\nonumber\\
K_5 &=&  K^E_5-2K^E_{18}\,,
\nonumber\\
K_6 &=&  K^E_6-K^E_{19}\,,
\nonumber\\
K_i &=&  K^E_i;\quad i=7,\ldots,14\,.
\end{eqnarray}

The extra subtractions needed can be derived from the divergences
of the $n$-flavour case.
We write
\be
\label{definf}
K^E_i = (e^c\mu)^{-2\epsilon}
\left(K^{Er}_i + k_i\frac{1}{16\pi^2 \epsilon}\right)\,,
\ee
with the dimension of space-time $d=4-2\epsilon$ and
\be
c = -\frac{1}{2}\left(\ln(4\pi) +\Gamma^\prime(1)+1\right)\,.
\ee
The equivalent subtractions needed for the $L_i$ can be found in
Ref.~\cite{GL2,BCE2}. We have derived the values of the $k_i$ from the
$n$-flavor results given in the appendix of Ref.~\cite{Knecht} after correcting
an obvious misprint and rewriting the terms in our minimal basis.
The $k_i$ are give in Tab.~\ref{tab:ki}.
\begin{table}[htbp]
  \begin{tabular}{|c|c|c|c|}
\hline
$i$ & \;\;\;\;\;\;\;\,$k_i$\;\;\;\;\;\;\;\, & $i$ & \;\;\;\;\;\;\;$k_i$ \;\;\;\;\;\;\;\\
\hline
1 & $0$             & 9  & $\frac{1}{8}$ \\
2 &$-\frac{1}{2}Z$ & 10 & $-\frac{1}{8}-\frac{3}{4}Z$ \\
3 & $0$             & 11 & $-\frac{1}{16}$ \\
4 & $-Z$            & 12 & $-\frac{1}{8}$ \\
5 & $\frac{3}{8}$   & 13 & $0$ \\
6 & $-\frac{3}{4}Z$ & 14 & $0$ \\
7 & $0$             & 18 & $-\frac{3}{8}$ \\
8 & $-\frac{1}{2}Z$ & 19 & $0$ \\
\hline
\end{tabular}
\caption{The values of the subtraction constants $k_i$ of PQ$\chi$PT.}
  \label{tab:ki}
\end{table}

\subsection{Propagators and LO masses}

The propagators for the supersymmetric formulation of PQ$\chi$PT can be found
 in Ref.~\cite{Sharpe2}. For calculational reasons, they have here
been translated from the Euclidean formalism into Minkowski space.
The charged propagators are given by~\cite{Sharpe2}
\begin{eqnarray}
-i\,G_{ij}^c (k) &=& 
\frac{\epsilon_j}{k^2 - M^2_{0,ij} + i\varepsilon}\quad (i \neq j)\,.
\label{propc}
\end{eqnarray}   
where $M^2_{0,ij}$ denotes the lowest order mass of the meson $\Phi_{ij}$ for
$i\ne j$ and the 
signature vector $\epsilon_j$ due to the graded structure is defined as
\begin{equation}
\epsilon_j=\left\{
  \begin{array}{cl}
    +1 &  \mathrm{for} \;\;\;j=1,\dots,6\\
    -1 & \mathrm{for}\;\;\;j=7,8,9\,.
  \end{array}
\right.
\end{equation}
For PQ$\chi$PT without electromagnetic interactions, the LO mass $M^2_{0,ij}$
is simply $\chi_{ij}\equiv (\chi_i+\chi_j)/2$. 
Electromagnetic interactions modify the
lowest order mass, and the new LO mass, which in the analytical
results below is denoted as $\chi_{e,ij}$, can be read off from the ${\cal
  O}(\Phi^2)$ terms of the lowest order Lagrangian. It is given by
\be
\label{LOmass}
\chi_{e,ij}=\chi_{ij}+\frac{2Ce^2}{F_0^2}\left( q_i-q_j\right)^2.
\ee
This is the mass that appears in the charged propagator.

The neutral propagators can have a double-pole structure and are more
complicated. However, since they are charge-neutral, the lowest order
propagator is unaffected by
the inclusion of electromagnetic corrections in the theory.
Explicit
expressions for the lowest order neutral propagators can be found
in~\cite{BDL2}. See also~\cite{Sharpe1}.

\section{Analytical results at NLO}
\label{Analytical}

The analytical expressions for the masses and decay constants are fairly
short, and very similar in form. Therefore it suffices to give the results for
the cases with $d_{\mathrm{sea}} = 3$ only.
They can also be downloaded from~\cite{website}.
Expressions for the cases with
$d_{\mathrm{sea}} = 2$ and $d_{\mathrm{sea}} = 1$ can easily be derived by
taking the appropriate limits, i.e. $\chi_5 \to \chi_4$ for 
$d_{\mathrm{sea}} = 2$ 
and $\chi_5, \chi_6 \to \chi_4$ for $d_{\mathrm{sea}} = 1$. It should be
noted, however, that for the degenerate cases, all sums are still over the
full set of indices, and furthermore, since we only take limits of the masses,
the charges $q_i$ are never affected by such limits. 

\subsection{Masses}

\begin{figure}[h!]
\begin{center}
\includegraphics[width=0.9\columnwidth]{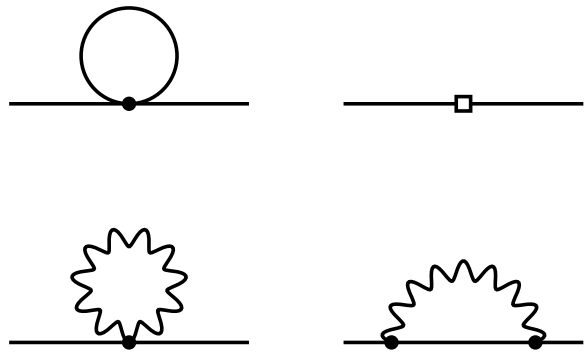}
\caption{The Feynman diagrams that 
contribute to $\Sigma_4(M_0^2,\chi_i)$ with electromagnetic corrections
included. A filled circle denotes a vertex from the ${\cal L}_2$ 
Lagrangian and an open square denotes a vertex from the ${\cal L}_4$ 
Lagrangian. A straight line is a pseudoscalar meson and a wiggly line is a 
photon.\label{massfig}}
\end{center}
\end{figure}

The corrections to the lowest order mass of a charged pseudoscalar meson is obtained
by calculating the self-energy corrections to the propagator in the
interacting theory, usually written in terms of the Fourier transform of the
two-point function
\begin{equation}
\label{propagator}
i\Delta(p)= \int d^4x\,e^{ip\cdot x} \langle \Omega \vert 
T[\Phi(x)_{ji}\Phi(0)_{ij}]
\vert \Omega \rangle,
\end{equation}
where $\Phi_{ij}=q_i \bar q_j$ denotes any of the off-diagonal mesons in 
the valence sector of PQ$\chi$PT, and $\Omega$ denotes the vacuum of 
the interacting theory. The propagator resums as a 
geometric series~\cite{Peskin}, giving
\begin{equation}
\label{PhysMass}
i\Delta(p)=\frac{i}{p^2-M_0^2-\Sigma(p^2,\chi_i)},
\end{equation}
where $M_0^2$ denotes the lowest order mass of the meson which is being 
considered, and $\chi_i$ in $\Sigma$ denotes the dependence of the 
self-energy on all the lowest order meson masses. The quantity 
$\Sigma(p^2,\chi_i)$ receives contributions from the 
one-particle-irreducible~(1PI) diagrams. The physical masses, 
are defined by the position of the pole in 
Eq.~(\ref{PhysMass}),
\begin{equation}
M_{\mathrm{phys}}^2=M_0^2+\Sigma(M_{\mathrm{phys}}^2,\chi_i),
\end{equation}
where the expression for the self-energy $\Sigma$ is written as a string of terms
denoting the 1PI diagrams of progressively higher order. The contributions
start at NLO, and thus
\ba
\Sigma(M_{\mathrm{phys}}^2,\chi_i) = 
\Sigma_4(M_{0}^2,\chi_i) +\mathcal{O}(p^6,e^2p^4),
\ea
Note that we have used the lowest order mass $M_{0}^2$ instead of $M_{\mathrm{phys}}^2$ 
in $\Sigma_4$ since the diagrams in that term are already of 
$\mathcal{O}(p^4,e^2p^2)$. The Feynman diagrams that 
contribute to $\Sigma_4(M_0^2,\chi_i)$ with electromagnetic corrections 
included are shown in Fig.~\ref{massfig}. 

We present the physical mass on the form
\begin{equation}
M_{\mathrm{phys}}^2 = \chi_{e,ij} + \frac{\delta^{(4)\mathrm{vs}}}{F_0^2} 
\:+\: \mathcal{O}(p^6,e^2p^4),
\label{delteq}
\end{equation}
where $\chi_{e,ij}=M_0^2$ is the lowest order mass, defined in 
Eq.~(\ref{LOmass}). 
The superscripts (v) and (s) indicate 
the values of 
$d_{\mathrm{val}}$ and $d_{\mathrm{sea}}$, respectively. It should also be
noted 
that the results are given in 
terms of the lowest order decay constant $F_0$ and the 
lowest order masses, since these are the fundamental inputs in 
PQ$\chi$PT. To the accuracy we are working with here they can be replaced by
the physical masses in the NLO correction.

The NLO contribution to the charged pseudoscalar meson mass with
electromagnetic corrections is for $d_{\mathrm{val}} = 1$ and
$d_{\mathrm{sea}} = 3$ found to be

\begin{eqnarray} 
\delta^{(4)13} &=& \left[ 48 L^r_6 -24 L^r_4\right] 
\chi_1\bar \chi_1  
+ \left[16 L^r_8 -8 L^r_5 \right] \chi_1^2 \nonumber \\ &&
- 48 e^2 F_0^2 Z_E L_4^r q_{12}^2 \bar \chi_1  
- 16 e^2 F_0^2 Z_E L_5^r q_{12}^2 \chi_1 \nonumber \\ &&
- e^2 F_0^2 \left[12 K_{1}^{Er} \!\!+\!\! 12 K_{2}^{Er} \!\!-\!\! 12 K_{7}^{Er} 
\!\!-\!\! 12 K_{8}^{Er} \right] \overline Q_2 \chi_1\nonumber \\ &&
-e^2 F_0^2\left[ 4 K_{5}^{Er}\!\!+\!\!4 K_{6}^{Er}\!\!-\!\! 4 K_{9}^{Er}
\!\!-\!\!4 K_{10}^{Er} \right] q_p^2 \chi_1\nonumber \\ &&
+ 12 e^2 F_0^2 K_{8}^{Er} q_{12}^2 \bar \chi_1\nonumber \\ &&
+ 8 e^2 F_0^2 \left[K_{10}^{Er}
\!\!+\!\! K_{11}^{Er}\right] q_{12}^2 \chi_1\nonumber \\ &&
- e^2 F_0^2\left[ 8 K_{18}^{Er} \!\!+\!\! 4 K_{19}^{Er}\right] q_1 q_2 \chi_1\nonumber \\ &&
-1/3 \bar A(\chi_m) R^m_{n11} \chi_1 
-1/3 \bar A(\chi_1) R^c_1 \chi_1\nonumber \\ &&
+  e^2 F_0^2 \bar A(\chi_1) q_{12}^2\nonumber \\ &&
+ 2 e^2 F_0^2 Z_E \bar A(\chi_{1s}) q_{12}^2\nonumber \\ &&
-1/3 \bar B(\chi_1,\chi_1,0)  R^d_1 \chi_1\nonumber \\ &&
+ 4 e^2 F_0^2 \bar B(\chi_{\gamma},\chi_1,\chi_1) 
q_{12}^2 \chi_1 \nonumber \\ &&
-4 e^2 F_0^2 \bar B_1(\chi_{\gamma},\chi_1,\chi_1) q_{12}^2 \chi_1. 
\end{eqnarray}

For $d_{\mathrm{val}} = 2$ and $d_{\mathrm{sea}} = 3$ one has
\begin{eqnarray} 
\delta^{(4)23} &=&
  \left[ 48 L_6^r - 24 L_4^r \right] \bar \chi_1 \chi_{13}
+ \left[ 16 L_8^r - 8 L_5^r \right] \chi_{13}^2\nonumber \\ &&
     -48 e^2 Z_E F_0^2 L_4^r q_{13}^2 \bar \chi_1
     -16 e^2 Z_E F_0^2 L_5^r q_{13}^2 \chi_{13}\nonumber \\ &&
 - e^2 F_0^2 \!\left[ 12 K^{Er}_1 \!\!+\!\! 12 K^{Er}_2 \!\!-\!\!12 K^{Er}_7 
\!\!-\!\! 12 K^{Er}_8 \right] \!\overline Q_2 \chi_{13}\nonumber \\ &&
- e^2 F_0^2 \left[ 4 K^{Er}_5 \!\!+ 4 K^{Er}_6 \right] q_p^2 \chi_{13}\nonumber \\ &&
+ e^2 F_0^2 \left[ 4 K^{Er}_9 \!\!+ 4 K^{Er}_{10} \right] q_p^2 \chi_p\nonumber \\ &&
+ 12 e^2 F_0^2 K^{Er}_8 q_{13}^2 \bar \chi_1\nonumber \\ &&
+ 8 e^2 F_0^2 \left[ K^{Er}_{10} \!\!+\!\! K^{Er}_{11} \right] q_{13}^2 \chi_{13}\nonumber \\ &&
- e^2 F_0^2 \left[ 8 K^{Er}_{18} + 4 K^{Er}_{19} \right] q_1 q_3
\chi_{13}\nonumber \\ &&
-1/3 \bar A(\chi_m) R^m_{n13} \chi_{13}
-1/3 \bar A(\chi_p) R^p_{q\pi\eta} \chi_{13}\nonumber \\ &&
 +  e^2 F_0^2 \bar A(\chi_{13}) q_{13}^2\nonumber \\ &&
+ 2 e^2 Z_E F_0^2 \bar A(\chi_{1s}) q_{1s} q_{13}\nonumber \\ &&
- 2 e^2 Z_E F_0^2 \bar A(\chi_{3s}) q_{3s} q_{13}\nonumber \\ &&
+ 4 e^2 F_0^2 \bar B(\chi_\gamma,\chi_{13},\chi_{13}) q_{13}^2 \chi_{13}\nonumber \\ &&
- 4 e^2 F_0^2 \bar B_1(\chi_\gamma,\chi_{13},\chi_{13}) q_{13}^2 \chi_{13}.
\end{eqnarray}

\subsection{Decay Constants}

\begin{figure}[h!]
\begin{center}
\includegraphics[width=0.9\columnwidth]{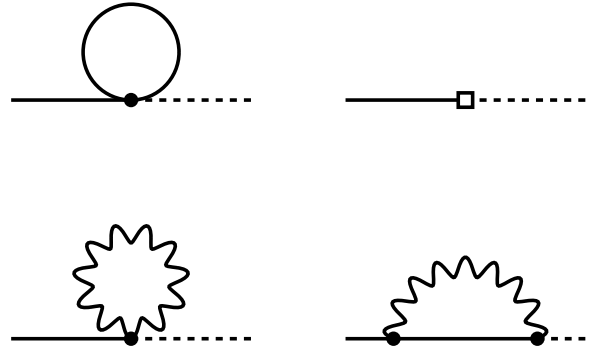}
\caption{The Feynman diagrams that contribute to the axial current operator at 
NLO. A filled circle denotes a vertex from the ${\cal L}_2$ Lagrangian and an 
open square denotes a vertex from the ${\cal L}_4$ Lagrangian. A straight 
line is a pseudoscalar meson, a wiggly line is a photon and a dashed 
line is an axial current.\label{decfig}}
\end{center}
\end{figure}

The decay constants $F_a$ of the pseudoscalar mesons are defined through
\begin{equation}
\langle 0| A_a^\mu(0) |\phi_a(p)\rangle = i\sqrt{2}\,p^\mu\,F_a,
\label{decdef}
\end{equation}
in terms of the axial current operator $A_a^\mu(0)$. In the following 
the flavor index $a$ has been suppressed for simplicity.
The Feynman diagrams that contribute to the axial current operator at 
NLO, are shown in Fig.~\ref{decfig}. 

Diagrams of ${\mathcal O}(p^4,e^2p^2)$
also contribute to Eq.~(\ref{decdef}) through
the wave function renormalization factor $\sqrt{{\mathcal Z}}$,
since the expression for the
decay constant of a meson is
\be
\frac{F_{\mathrm{phys}}}{\sqrt{{\mathcal Z}}} = 
F_0 + F_4(M_{\mathrm{phys}}^2,\chi_i) + {\mathcal O}(p^6,e^2p^4).
\label{deceq}
\ee
In
Eq.~(\ref{deceq}), the 
subscripts of the matrix elements $F$ indicate the chiral order and the 
lowest order contribution $F_2$ has been identified with $F_0$. 
The wave function renormalization is given in terms of the self-energy 
diagrams by
\begin{eqnarray}
{\mathcal Z}^{-1} &\equiv&
1 - \left.\frac{\partial\Sigma(p^2,\chi_i)}
{\partial p^2}\right|_{M_{\mathrm{phys}}^2}
\end{eqnarray}
which becomes, when expanded such that all contributions up to 
${\mathcal O}(p^4,e^2p^2)$ are taken into account,
\begin{eqnarray}
\sqrt{{\mathcal Z}} &=& 1 \:+\: \frac{\Sigma'}{2} 
\:+\: \cdots \\
\Sigma' &=& 
\left.\frac{\partial\Sigma_4(p^2,\chi_i)}
{\partial p^2}\right|_{M_{\mathrm{phys}}^2} \,.
\end{eqnarray}
The quantity $\Sigma_4$ denotes the one-particle-irreducible diagrams to
${\cal O}(p^4,e^2p^2)$.
Combining these expressions, the decay 
constant at NLO is then
\begin{eqnarray}
F_{\mathrm{phys}} &=& F_0 \:+\:F_4(\chi_i)
\nonumber \\ 
&+& F_0 \left.\frac{\partial\Sigma_4(p^2,\chi_i)}
{2\,\partial p^2}\right|_{M_0^2} +{\mathcal O}(p^6,e^2p^4)\,.
 \end{eqnarray}
Again it is sufficient to use the lowest order mass $M_{0}^2$
in $\Sigma_4$ since the diagrams in that term are
already of $\mathcal{O}(p^4,e^2p^2)$.
The analytical results for the decay constant are below given in the form
\begin{equation}
F_{\mathrm{phys}} = F_0 \left[ 1 + \frac{f^{(4)\mathrm{vs}}}{F_0^2} 
+ \mathcal{O}(p^6,e^2p^4) \right],
\label{delteqF}
\end{equation}
As for the meson masses, the 
superscripts (v) and (s) indicate the values of $d_{\mathrm{val}}$ and 
$d_{\mathrm{sea}}$, respectively.

The NLO contribution to the decay constant for a charged pseudoscalar 
meson with
electromagnetic corrections is for $d_{\mathrm{val}} = 1$ and
$d_{\mathrm{sea}} = 3$ found to be

\begin{eqnarray} 
f^{(4)13} &=&  
+ 12 L_4^r \bar \chi_1
+ 4 L_5^r \chi_1\nonumber \\ &&
+ 6  e^2 F_0^2 \left[K^{Er}_1 + K^{Er}_2 \right] \overline Q_2\nonumber \\ &&
+ 2 e^2 F_0^2 \left[ K^{Er}_5 + K^{Er}_6 \right] q_p^2\nonumber \\ &&
+ 2  e^2 F_0^2 K^{Er}_{12} q_{12}^2\nonumber \\ &&
+  e^2 F_0^2 \left[ 4 K^{Er}_{18} + 2 K^{Er}_{19} \right] q_1 q_2\nonumber \\ &&
+1/4 \bar A(\chi_{e,ps}) \nonumber \\ && 
+ 2 e^2 F_0^2 \bar B^\prime(\chi_\gamma,\chi_1,\chi_1) q_{12}^2 \chi_1\nonumber \\ &&
- e^2 F_0^2 \bar B_1(\chi_\gamma,\chi_1,\chi_1) q_{12}^2\nonumber \\ &&
-2 e^2 F_0^2 \bar B_1^\prime(\chi_\gamma,\chi_1,\chi_1) q_{12}^2 \chi_1.
\end{eqnarray}

For $d_{\mathrm{val}} = 2$ and $d_{\mathrm{sea}} = 3$ the result is
\begin{eqnarray} 
f^{(4)23} &=& =
+ 12 L_4^r  \bar \chi_1
+ 4 L_5^r  \chi_{13}\nonumber \\ &&
+ 6 e^2 F_0^2 \left[ K^{Er}_1 + K^{Er}_2 \right]\overline Q_2\nonumber \\ &&
+ 2 e^2 F_0^2 \left[ K^{Er}_5 + K^{Er}_6 \right] q_p^2\nonumber \\ &&
+ 2 e^2 F_0^2 K^{Er}_{12} q_{13}^2\nonumber \\ &&
+ e^2 F_0^2 \left[ 4 K^{Er}_{18} + 2 K^{Er}_{19} \right] q_1 q_3\nonumber \\ &&
- 1/12 \bar A(\chi_m) R^v_{mn13}\nonumber \\ &&
+ \bar A(\chi_p) \left[1/6 R^p_{q\pi\eta} - 1/12 R^c_p \right]\nonumber \\ &&
+1/4 \bar A(\chi_{e,ps})\nonumber \\ &&
-1/12 \bar B(\chi_p,\chi_p,0) R^d_p\nonumber \\ &&
+ 2 e^2 F_0^2 \bar B^\prime(\chi_\gamma,\chi_{13},\chi_{13}) q_{13}^2 \chi_{13}\nonumber \\ &&
- e^2 F_0^2 \bar B_1(\chi_\gamma,\chi_{13},\chi_{13}) q_{13}^2\nonumber \\ &&
- 2 e^2 F_0^2 \bar B_1^\prime(\chi_\gamma,\chi_{13},\chi_{13}) q_{13}^2 \chi_{13}.
\end{eqnarray}
The term containing $R^v_{mn13}$ is somewhat tricky to take the limit
to the simpler mass cases. The form needed for the simpler cases
can be found in Ref.~\cite{BL1} or in~\cite{website}.

\section{Discussion of the analytical results}
\label{analytical}

Our analytical results are finite. The renormalization obtained from
the $n$-flavour divergences using the arguments presented above and in
our earlier work,
did cancel those from the loop diagrams. In addition, they agree with earlier
PQ$\chi$PT results when the electromagnetic parts are removed as well as with
the known results for electromagnetic corrections when removing the
partial quenching. We have used the definition of the decay constant with the
axial current. This definition has an infrared divergence as can be seen also
in our result. We have regulated that divergence with a photon mass
$\chi_\gamma$. This is the definition which was used in Ref.~\cite{Neufeld}
as well. How to relate this to measurable quantities can be found in
Ref.~\cite{Cirigliano}.

Which combinations of the new
LECs can now be determined from lattice calculations?
In the masses 5 independent combinations appear:
\begin{eqnarray}
Y_1&=& K^{Er}_1+K^{Er}_2-K^{Er}_7-K^{Er}_8\,,
\nonumber\\
Y_2&=& K^{Er}_9+K^{Er}_{10}\,,
\nonumber\\
Y_3&=&-K^{Er}_5-K^{Er}_6+2K^{Er}_{10}+2K^{Er}_{11}\,,
\nonumber\\
Y_4&=&2K^{Er}_5+2K^{Er}_6+2K^{Er}_{18}+K^{Er}_{19}\,,
\nonumber\\
Y_5 &=&K^{Er}_8
\end{eqnarray}
These can be determined by varying the charges
and quark masses separately. It should be noted that the
sea quark charges only have a dependence via $Y_1\overline{Q}_2\chi_{13}$
with undetermined LECs.

The decay constants depend on the combinations
\begin{eqnarray}
Y_6&=& K^{Er}_1+K^{Er}_2\,,
\nonumber\\
Y_7&=& K^{Er}_5+ K^{Er}_6+ K^{Er}_{12}\,,
\end{eqnarray}
as well as on $Y_4$.  It should be noted that the
sea quark charges only have a dependence via $Y_6\overline{Q}_2$
with undetermined LECs.

There is in addition dependence on the sea-quark charges in the chiral
logarithms, but this dependence is predicted at NLO.

The individual masses and decay constants depend on the sea quark charges.
But since the overall dependence on the sea quark charges appearing
with unknown LECs is simple we can easily make combinations where
this disappears. We use here the notation
\begin{equation}
M^2(\chi_1,\chi_3,q_1,q_3)
\end{equation}
to denote the mass of the meson with valence masses $\chi_1$ and $\chi_2$
and valence charges $q_1$ and $q_3$.
The quantity
\begin{eqnarray}
\label{defDM}
\Delta M^2 &=& M^2(\chi_1,\chi_3,q_1,q_3)
- M^2(\chi_1,\chi_3,q_3,q_3)
\nonumber\\
&&- M^2(\chi_1,\chi_1,q_1,q_3)
+ M^2(\chi_1,\chi_1,q_3,q_3)
\end{eqnarray}
is especially useful. Only the electromagnetic corrections
survive and the only dependence on the sea-quark charges is in some of the
chiral logarithms. Since these contributions are independent of the
LECs, they can be subtracted before making fits with lattice simulations, and
hence do not present any problem in this respect.
The quantity in Eq.~(\ref{defDM}) is also directly relevant for 
the violation of
Dashen's theorem~\cite{Dashen,BijnensDashen},
\begin{equation}
\Delta M^2_D = (m^2_{K^+}-m^2_{K^0})- (m^2_{\pi^+}-m^2_{\pi^0})\,.
\end{equation}
Dashen's theorem states that the electromagnetic part of $\Delta M^2_D$
vanishes. $\Delta M^2$ becomes the electromagnetic part of $\Delta M^2_D$
up to some very small electromagnetic corrrections to the $\pi^0$ mass in the
isospin limit.

Similarly, differences of decay constants of particles containg valence quarks
with the same charges have no dependence on the sea quark charges with
unknown LECs. In particular this true for the difference of the pion
and kaon decay constant. We define
\begin{equation}
F(\chi_1,\chi_3,q_1,q_3,\{q_{sea}\})
\end{equation}
to be the decay constant of a meson with valence masses and charges as for
$M^2$ above and sea-quark charges $\{q_{sea}\}$.
The quantity
\begin{eqnarray}
\label{defDF}
\Delta F &=&\bigg[F(\chi_1,\chi_3,q_1,q_3, \{q_{sea}\})-
 F(\chi_1,\chi_3,0,0, \{0\})
\nonumber\\
&&-F(\chi_1,\chi_1,q_1,q_3, \{q_{sea}\})+
 F(\chi_1,\chi_1,0,0, \{0\})\bigg]
\nonumber\\
&&/F_0\,,
\end{eqnarray}
is an example of this.
It gives the relative electromagnetic corrections to the difference of kaon
and pion decay constants. In fact, $\Delta F$ is independent of all
the $K^{Er}_i$.

\section{Numerical results}
\label{numerical}

The whole purpose of this work is that our formulas can be used by the
lattice QCD community to perform their fits. We therefore only present some
representative numerical results.
For the $L_i^r$ we use the values determined in the NNLO order fit of
Ref.~\cite{ABT2}, called fit 10. For the extra electromagnetic parameters we use the estimates
of Ref.~\cite{BP}. There are four combinations of the $K^{Er}_i$ estimated
there. We simply choose a series of $K_i^{Er}$ values that reproduces
the combinations estimated there and set all others to zero. The 
nonzero values we have
chosen for illustration are
\begin{eqnarray}
  \label{valuesKi}
K^{Er}_5&=& 2.85\cdot 10^{-3}\,,\quad K^{Er}_9 = 1.3\cdot 10^{-3}\,,
\nonumber\\
K^{Er}_{10}&=& 4.0\cdot 10^{-3}\,,\quad K^{Er}_{11} = -1.25\cdot 10^{-3}\,,
\nonumber\\
C&=& Z_E F_0^4 = 4.2\cdot 10^{-5}~\mbox{GeV}^5\,,
\end{eqnarray}
at a subtraction scale $\mu=770$~MeV. Earlier estimates of $C$ are
in Refs.~\cite{Bardeen,EGPR}.

The numerics we present here uses $q_1 = q_4 =2/3$ and
$q_2=q_3=q_5=q_6=-1/3$ and a value of $e$ determined from the measured
fine structure constant $\alpha$. We also only quote the electromagnetic
part by subtracting the same result with $e=0$.

The lowest order correction to the meson masses vanishes for those
with zero total charge. For charged mesons it is equal to
\begin{equation}
  \label{eqLO}
M^2_{LO} = 1.00\cdot 10^{-3}~\mbox{GeV}^{2}\,.
\end{equation}
This should be compared to the physical mass difference
\begin{equation}
  \label{eqphys}
m^2_{\pi^+}-m^2_{\pi^0} = 1.3\cdot 10^{-3}~\mbox{GeV}^{2}\,.  
\end{equation}
There is no electromagnetic correction to the decay constants at lowest order.

Below we use for convenience the terminology $\pi$ for a meson
with both valence masses equal to $\chi_1$ and $K$ for a meson with
valence masses equal to $\chi_1$ and $\chi_3$ respectively.
The charge label is $+$ for valence quark charges $2/3$ and $-1/3$
and $0$ for valence quark charges $-1/3$ and $-1/3$, i.e. electrically neutral.

We first quote the electromagnetic corrections for $\chi_1=\chi_4=\chi_5$ 
and $\chi_3=\chi_6$
with $\sqrt{\chi_1}=135$~MeV and $\sqrt{\chi_{13}}=495$~MeV.
\begin{eqnarray}
\label{resultM}
M^2_{\pi^+NLO}&=&0.45\cdot 10^{-3}~\mbox{GeV}^{2}\,,
\nonumber\\
M^2_{K^+NLO}&=&1.52\cdot 10^{-3}~\mbox{GeV}^{2}\,,
\nonumber\\
M^2_{\pi^0NLO}&=&-2\cdot 10^{-7}~\mbox{GeV}^{2}\,,  
\nonumber\\
M^2_{K^0NLO}&=&-3\cdot 10^{-6}~\mbox{GeV}^{2}\,.  
\end{eqnarray}
This leads to a value of
\begin{equation}
\Delta M^2 = 1.07  \cdot 10^{-3}~\mbox{GeV}^{2}\,.
\end{equation}
In agreement with the large violation of Dashen's theorem seen in
Ref.~\cite{BP} since we used their estimate for the constants and similar
values for the other inputs.
The electromagnetic corrections for the decay constants with a photon mass
$\chi_\gamma = (10~\mbox{MeV})^2$ are
\begin{eqnarray}
  \label{eq:resultF}
F_{\pi^+NLO}/F_0 &=& 0.0039\,.
\nonumber\\
F_{K^+NLO}/F_0 &=& 0.0056\,,
\end{eqnarray}
leading to
\begin{equation}
  \label{eq:DF}
  \Delta F = 0.0017\,.
\end{equation}

\begin{figure}[htbp]
\includegraphics[width=\columnwidth]{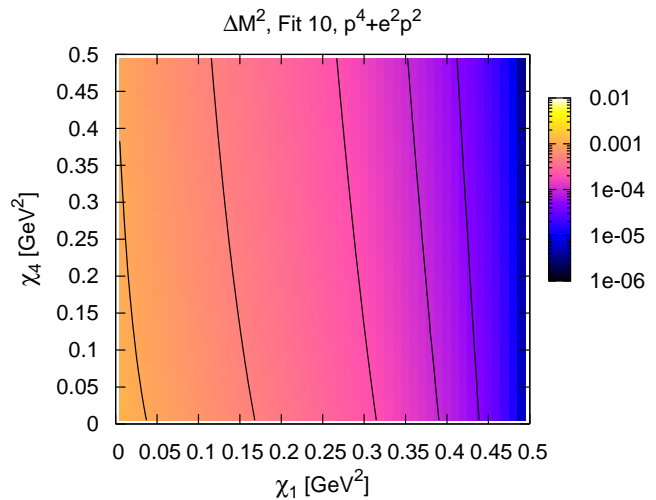}
  \caption{The quantity $\Delta M^2$ of Eq.~\ref{defDM}, the difference of
electromagnetic contributions to
meson masses between kaons and pions as a function of the
input lowest order masses $\chi_1$ and $\chi_4$. The scale is logarithmic
and contour lines are drawn
at $\Delta M^2=0.00005,0.0001,0.0002,0.0005,0.001$. }
  \label{fig:DM}
\end{figure}

\begin{figure}[tbp]
\includegraphics[width=\columnwidth]{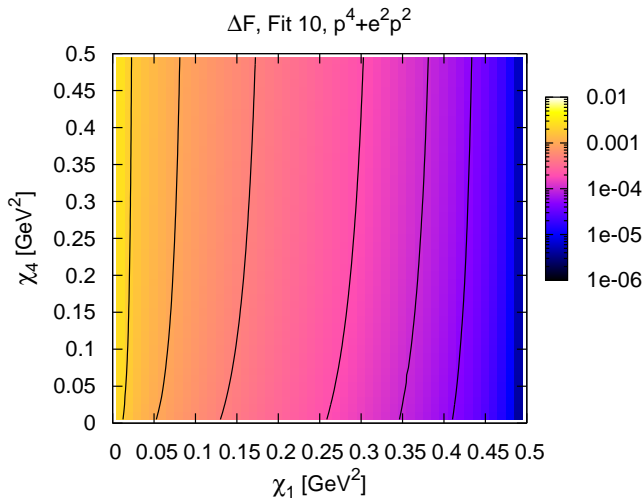}
  \caption{The quantity $\Delta F$ of Eq.~\ref{defDF}, the relative
difference of
electromagnetic contributions to meson decay constants between kaons and pions
as a function of the
input lowest order masses $\chi_1$ and $\chi_4$. The scale is logarithmic
and contour lines are drawn
at $\Delta F= 0.00005,0.0001,0.0002,0.0005,0.001,0.002$.}
  \label{fig:DF}
\end{figure}

The above results are for the unquenched case. To show the effects of partial
quenching we plot the quantities $\Delta M^2$ and $\Delta F$ with input
values as above and $\chi_4=\chi_5$ and $\chi_3=\chi_6=0.5~\mbox{GeV}^2$
as a function of $\chi_1$ and $\chi_4$ in Figs.~\ref{fig:DM} and \ref{fig:DF}.

\begin{figure}[tbp]
  \includegraphics[width=\columnwidth]{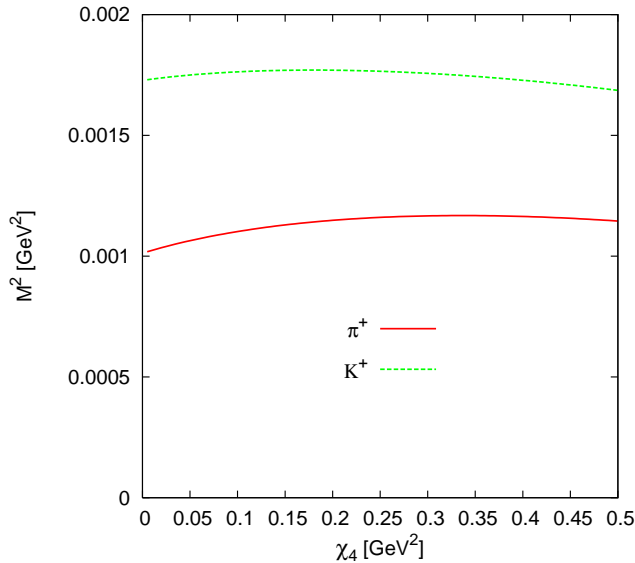}
  \caption{The NLO electromagnetic correction as a function of the
sea quark mass $\chi_4$ for the case of $\pi^+$ and $K^+$.
See text for the other inputs.}
  \label{fig:piK}
\end{figure}

\section{Conclusions}
\label{conclusions}

In this paper we have shown how to include electromagnetic corrections in
partially quenched Chiral Perturbation Theory. We have then used this formalism
to compute the electromagnetic corrections to masses and decay constants of
the charged or off-diagonal mesons to NLO in PQ$\chi$PT. We also presented
some illustrative numerical results.

We have shown that for several phenomenologically interesting quantities
the relevant LECs can be computed using quenched photons, i.e. they
can be computed with the photons only coupling to the valence quarks.

The dependence on the sea quark mass is rather small in these differences.
It cancels to a large extent.
In Fig.~\ref{fig:piK} we show the electromagnetic contribution to the
squared mass of the pion and kaon as a function of 
$\chi_4=\chi_5$ with $\chi_1=0.1~\mbox{GeV}^2$
and the other inputs as above.

\acknowledgments

\texttt{FORM 3.0} was used heavily in these calculations~\cite{FORM}.
This work is supported by the European Union RTN network,
Contract No. 
MRTN-CT-2006-035482  (FLAVIAnet) and by 
the European Community-Research Infrastructure
Activity Contract No. RII3-CT-2004-506078 (HadronPhysics).

\end{document}